\documentclass[preprint,nofootinbib]{revtex4}%
\usepackage{amssymb}
\usepackage{amsfonts}
\usepackage{amsmath}
\usepackage{amsmath}
\usepackage{graphicx}
\usepackage[usenames]{color}%
\providecommand{\U}[1]{\protect\rule{.1in}{.1in}}

\providecommand{\U}[1]{\protect\rule{.1in}{.1in}}
\definecolor{blue}{rgb}{0,0,1}

\definecolor{red}{rgb}{1,0,0}

\begin{document}
\title{Primary scalar hair in Gauss-Bonnet black holes with Thurston horizons}
\author{Luis Guajardo$^1$ and Julio Oliva$^2$}

\affiliation{$^1$Instituto de Investigaci\'on Interdisciplinaria, Vicerrector\'ia Acad\'emica, Universidad de Talca, 3465548 Talca, Chile.}

\affiliation{$^2$Departamento de F\'{\i}sica, Universidad de Concepci\'{o}n, Casilla, 160-C,
Concepci\'{o}n, Chile.}

\begin{abstract}
In this work, we construct novel asymptotically locally AdS$_5$ black hole solutions of Einstein-Gauss-Bonnet theory at the Chern-Simons point, supported by a scalar field that generates a primary hair. The strength of the scalar field is governed by an independent integration constant; when this constant vanishes, the spacetime reduces to a black hole geometry devoid of hair. The existence of these solutions is intrinsically tied to the horizon metric, which is modeled by three non-trivial Thurston geometries: Nil, Solv, and $SL(2,\mathbb{R})$. The quadratic part of the scalar field action corresponds to a conformally coupled scalar in five dimensions -an invariance of the matter sector that is explicitly broken by the introduction of a quartic self-interaction. These black holes are characterized by two distinct parameters: the horizon radius and the temperature. Notably, there exists a straight line in this parameter space along which the horizon geometry exhibits enhanced isometries, corresponding to solutions previously reported in {\it{JHEP 02, 014 (2014)}}. Away from this line, for a fixed horizon radius and temperatures above or below a critical value, the metric's isometries undergo spontaneous breaking. Employing the Regge-Teitelboim approach, we compute the mass and entropy of these solutions, both of which vanish. Despite this, only one of the integration constants can be interpreted as hair, as the other modifies the local geometry at the conformal boundary. Finally, for Solv horizon geometries, we extend these hairy solutions to six dimensions.
\end{abstract}
\maketitle

\section{Introduction \label{sec:Intro}}

As compared with other astrophysical objects, black holes are very simple. In General Relativity (GR) in arbitrary dimensions they are described by few numbers corresponding to global charges like the mass and angular momenta, and other labels that can be used to describe the topology of the horizon \cite{Horowitz:2012nnc}. In presence of matter fields many no-hair results constrain the existence of black holes with non-trivial scalars (for a review, see \cite{Herdeiro:2015waa} and references therein and thereof), which nevertheless can be by-passed in sensible setups as it occurs with self-interacting scalar fields (see e.g. \cite{sc1}-\cite{sc4}) and field with conformal couplings \cite{Bocharova:1970skc}-\cite{Ortaggio:2024afv}, which will play a role in this paper.

On the other hand, going beyond Einstein's gravity, considering higher curvature terms in its own right for specific values of the coupling constants, may lead to scenarios with enlarged invariance in the gravity sector, beyond the local Lorentz $SO(d,1)$ symmetry. This is the case, for example, when the Einstein-Hilbert Lagrangian with a cosmological constant is supplemented with the Gauss-Bonnet term and there is a precise relation between the Gauss-Bonnet coupling $\alpha$ and the bared cosmological term $\Lambda$ in the action. Indeed, in dimension five when $\Lambda\alpha=-3/4$, the theory enjoys an enlarged symmetry given by the action of the local (Anti-)de Sitter group, which acts on a gauge connection constructed out from the vielbein and the spin connection, and the action can be written as a Chern-Simons action \cite{Chamseddine:1989nu} (see also \cite{Hassaine:2016amq} for a comprehensive list of references). It is known that in vacuum, in such special point, the space of solutions is also enlarged and the theory admits exact, rotating spacetimes \cite{Anabalon:2009kq,Cvetic:2016sow,Tapia:2024xcp}, wormholes \cite{Dotti:2006cp,Dotti:2007az}, dimensionally continued black holes\footnote{For topological black holes in GR as well as black holes with Thurson horizons see \cite{Lemos:1994xp,Mann:1996gj,Vanzo:1997gw,Brill:1997mf} and \cite{Cadeau:2000tj,Hassaine:2015ifa}.
} \cite{Banados:1993ur} and even asymptotically Lifshitz black holes \cite{Dehghani:2010kd}. When matter fields are included, both in the context of Chern-Simons supergravity \cite{Giribet:2014hpa,Andrianopoli:2021qli} and in non-supersymmetric setups \cite{BravoGaete:2013acu,Correa:2013bza}, the space of solutions is enlarged as well. In the latter families of solutions, when the matter fields correspond to a scalar field, the fully backreacting solutions are characterized by a single integration constant, and the black holes obtained are therefore dressed with a scalar hair of the secondary type, in consequence when the scalar field vanishes the spacetime reduces to the constant curvature background. The main aim of the present work is to present a novel family of black hole solution of Einstein-Gauss-Bonnet theory in five dimensions, at the Chern-Simons point, which is supported by a fully backreacting scalar field standing for a primary scalar hair, namely, the static black hole solution is characterized by two independent integration constants, one having a gravitational origin and the other coming from a scalar field. In the absence of the latter, the spacetime metric also describes a black hole. This effect is achieved by considering homogeneous, but not maximally symmetric horizons, modeled by Thurston geometries of the Nil, Solv and $SL(2,\mathbb{R})$ type. There is a 
particular limit for which our new solutions map to black holes with planar horizons, and in such case the spacetime metric consistently reduces to the one found in \cite{Correa:2013bza}. The new solutions presented in this paper can be interpreted as a hairy extension of the Dimensionally Continued Black Holes of \cite{Banados:1993ur}, which branches out continuously from the latter due to the presence of an independent integration constant and a different geometry at the horizon, that propagates to the conformal boundary of the asymptotically, locally AdS black holes.

\section{Black hole with primary hair \label{sec:BH}}
The action principle of our model is given by 
\begin{align}
\label{eq:action}
    S[g_{\mu\nu}, \phi] = \int d^5x\sqrt{-g}\left[ \dfrac{1}{2\kappa}\left( R - 2\Lambda + \alpha \mathcal{G} \right)- \dfrac{1}{2}\nabla_{\mu}\phi\nabla^{\mu}\phi - \dfrac{3}{32}R\phi^2 - \nu \phi^4 \right]\,,
\end{align}
where $\mathcal{G} = R^2 - 4R_{\alpha\beta}R^{\alpha\beta} + R_{\alpha\beta\gamma\delta}R^{\alpha\beta\gamma\delta}$ is the Gauss-Bonnet density. As it is well-known, despite having quadratic terms in the curvature, the field equations are of second order \cite{Lovelock:1971yv}. Indeed, the field equations read
\begin{subequations}
\begin{align}
\label{eq:feq} & G_{\mu\nu} + \Lambda g_{\mu\nu} + \alpha K_{\mu\nu} = \kappa T_{\mu\nu}\,,\\
    \label{eq:kleingordon} & \square \phi = \dfrac{3}{16}R\phi + 4\nu \phi^3\,,
\end{align}
\end{subequations}
where
\begin{align}
K_{\mu\nu} &= 2RR_{\mu\nu} - 4R_{\mu \sigma \nu \rho}R^{\sigma \rho} + 2R_{\mu \sigma \rho \tau}R_{\nu}^{\ \sigma \rho \tau} - 4R_{\mu \sigma}R_{\nu}^{\ \sigma} - \frac{1}{2}g_{\mu\nu} \mathcal{G} \,,
\end{align}
and the stress-energy tensor is given by
\begin{align}
\label{eq:tmunu}
    T_{\mu\nu} &= \nabla_{\mu}\phi\nabla_{\nu}\phi - g_{\mu\nu}\left(\frac{1}{2}\nabla_{\sigma}\phi\nabla^{\sigma}\phi + \nu \phi^4 \right) + \dfrac{3}{16} (g_{\mu\nu}\square - \nabla_{\mu}\nabla_{\nu} + G_{\mu\nu} )\phi^2\,.
\end{align}

We will focus on the Chern-Simons-AdS case, for which $\Lambda \alpha = -3/4$, where we have set the radius of the unique maximally symmetric AdS$_5$ solution to $\ell=1$, and we look for black holes over a homogeneous ansatz,
\begin{align}
    ds^2 = -f(r)~dt^2 + \dfrac{dr^2}{f(r)} + r^2d\Omega_{3}^2,,
\end{align}
with $d\Omega_{3}$ standing for the line element of any three-dimensional Thurston geometry in the family Nil, Solv and $SL(2\mathbb{R})$. The solution stems from a combination of the field equations, leading to a relation between the metric function $f(r)$ and the scalar field $\phi 
= \phi(r)$, in the form
\begin{align}
    \alpha f'' - \dfrac{1}{2} + \dfrac{3}{32} \kappa \phi^2=0\,,
\end{align}
which will lead to a simple exact solution.

The five-dimensional scalar field is generically given by
\begin{align}
\label{eq:scalar_field}
    \phi(r)= \dfrac{A}{r^{3/2}}\,,
\end{align}
where $A$ is an integration constant, and the parameter $\nu$, measuring the strength of the self-interacting potential, is related to $\alpha$ by
\begin{align}
\label{eq:nu} \nu = -\dfrac{27\kappa}{2048\alpha}\,.    
\end{align}

The model \eqref{eq:action} is effectively described by a single parameter, $\alpha$, which supports the Gauss-Bonnet density and intertwines with the cosmological term and with the coupling of the quartic potential. The latter, being different to $\phi^{\frac{2D}{D-2}}$ with $D=5$, explicitly breaks conformal invariance. In this sense, the situation is similar to the four-dimensional scalar-tensor family constructed in reference \cite{Fernandes:2021dsb}. Although the model studied here does not have a conformally invariant scalar equation, it is still parametrized by a single number, and it is also invariant under the $\mathbb{Z}_2$ symmetry, $\phi \mapsto -\phi$.

In what follows, we will present the metrics of the solutions for the different horizon geometries, in an explicit manner.

\subsection{Hairy Nil horizons}
For the Nil horizon geometry, the hairy black hole metric reads
{\small{
\begin{equation}
\label{eq:nil_metric}
ds^2=-\left(\dfrac{r^2}{4\alpha} - \mu - \dfrac{3\kappa A^2}{64\alpha r}\right)dt^2+\frac{dr^2}{\left(\dfrac{r^2}{4\alpha} - \mu - \dfrac{3\kappa A^2}{64\alpha r}\right)}+r^2\left[dx^2 + dy^2 + (dz - 2\sqrt{\mu}xdy)^2 \right]\ ,
\end{equation}
}
The Killing vectors of the spacelike section can be written as
\begin{align}\label{KVNil}
    \zeta_1 = \partial_x + 2\sqrt{\mu}y\partial_z\,,\quad \zeta_2 = \partial_z\,,\quad  \zeta_3 =\partial_y\,,\quad \zeta_4=y\partial_x - x\partial_y +2\sqrt{\mu} \left(\frac{y^2}{2} -\frac{x^2}{2}\right)\partial_z\,, 
\end{align}
satisfying the following non-vanishing commutation relations
\begin{align}
    \label{eq:nil_lie_structure}
    [\zeta_1, \zeta_3] = -2\sqrt{\mu}\zeta_2\,,\quad [\zeta_1,\zeta_4]=-\zeta_3\,,\quad [\zeta_3,\zeta_4] = \zeta_1\,.
\end{align}
Notice that the direction parameterized by the coordinate $z$ can be periodically identified, without spoiling the global definition of the Killing fields in \eqref{KVNil}.

\subsection{Hairy Solv horizons}

The metric for the Solv horizons yields
\begin{equation}
\label{eq:solv_metric}
ds^2=-\left(\dfrac{r^2}{4\alpha} - \mu - \dfrac{3\kappa A^2}{64\alpha r}\right)dt^2+\frac{dr^2}{\left(\dfrac{r^2}{4\alpha} - \mu - \dfrac{3\kappa A^2}{64\alpha r}\right)}+r^2\left[e^{2\sqrt{\mu}z}dx^2 + e^{-2\sqrt{\mu}z}dy^2 + dz^2 \right]\ ,
\end{equation}
The Killing vectors of the spacelike section in this case read
\begin{align}
    \zeta_1 = -\sqrt{\mu}x\partial_x + \sqrt{\mu}y\partial_y + \partial_z \,,\quad \zeta_2 = \partial_y\,,\quad \zeta_3 = \partial_x\,, 
\end{align}
and have the following non-vanishing commutators
\begin{align}\label{eq:solv_lie_structure}
    [\zeta_1, \zeta_2] = -\sqrt{\mu}\zeta_2\,,\quad [\zeta_1,\zeta_3] = \sqrt{\mu}\zeta_3\,.
\end{align}
For this geometry, the coordinate $x$ or $y$ can be compactified, which lead to the breaking of the global definition of the isometry generated by $\zeta_1$.

\subsection{Hairy $SL(2,\mathbb{R})$ horizons}
Finally, the new hairy black hole solution of Einstein-Gauss-Bonnet theory, with a horizon modeled by the $SL(2,\mathbb{R})$ Thurston geometry reads
\begin{equation}
\label{eq:sl2_metric}
ds^2=-\left(\dfrac{r^2}{4\alpha} - \mu - \dfrac{3\kappa A^2}{64\alpha r}\right)dt^2+\frac{dr^2}{\left(\dfrac{r^2}{4\alpha} - \mu - \dfrac{3\kappa A^2}{64\alpha r}\right)}+r^2\left[\frac{R_0^2}{x^2}\left(dx^2+dy^2\right)+\left(dz-R_0\frac{dy}{x}\right)^2\right]\ ,
\end{equation}
where the scale $R_0$
is fixed as
\begin{equation}
R_0=\sqrt{\frac{5}{4\mu}}
\end{equation}
and again the scalar field is given by equation \eqref{eq:scalar_field}. In this case, the Killing fields of the spacelike section are given by
\begin{equation}
\zeta_1=R_0^{-1}\partial_y, \ \zeta_2=\partial_z,\ \zeta_3=-R_0^{-1}\left(x\partial_x+y\partial_y\right),\ \zeta_4=R_0^{-1}\left(-xy\partial_x+\frac{y^2-x^2}{2}\partial_y+x\partial_z\right). 
\end{equation}
fulfilling the following, non-vanishing commutators
\begin{align}\label{eq:sl2_lie_structure}
    [\zeta_1,\zeta_3] = -R_0^{-1}\zeta_1\,,\quad [\zeta_1,\zeta_4]=R_0^{-1}\zeta_3\,,\quad [\zeta_3,\zeta_4] = -R_0^{-1}\zeta_4\,.
\end{align}
The coordinates $(x,y,z)$ of the horizon of \eqref{eq:sl2_metric} correspond to the $SL(2,\mathbb{R})$ geometry written as a fibration over the two-dimensional half-plane. These coordinates are not suitable for taking the planar limit $\mu=0$. In order to do so, it is useful to rewrite the horizon manifold as a fibration over the two-dimensional Poincare disk, which is achieved by the following coordinate transformation
\begin{equation}
x=\frac{4R_0^2-X^2-Y^2}{(X-2R_0)^2+Y^2}, \ y=-\frac{4R_0Y}{(X-2R_0)^2+Y^2}, \ z=Z-Y\ .
\end{equation}
In the new coordinates $(t,r,X,Y,Z)$, the solution \eqref{eq:sl2_metric} is rewritten as
{\small
\begin{equation}
\label{eq:sl2_metricPD}
ds^2=-f\left(r\right)dt^2+\frac{dr^2}{f(r)}+r^2\left[\frac{dX^2+dY^2}{\left(1-\frac{1}{4R_0^2}\left(X^2+Y^2\right)\right)^2}+\left(dZ+B(X,Y)dX+C(X,Y)dY\right)^2\right]\ ,
\end{equation}
}
with the $f(r)$ function given in \eqref{eq:sl2_metric} and the functions $B(X,Y)$ and $C(X,Y)$ defined as
\begin{align}
B(X,Y)&=\frac{8YR_0^2(X-2R_0)}{((X-2R_0)^2+Y^2)(X^2-4R_0^2+Y^2)}\\
C(X,Y)&=-\frac{(2(Y-X)R_0+Y^2+X^2)(X^2+Y^2-2(Y+X)R_0)}{((X-2R_0)^2+Y^2)(X^2-4R_0^2+Y^2)}\ .
\end{align}
Notice that in these coordinates one can take the limit $\mu\rightarrow 0$, namely $R_0\rightarrow \infty$ straightforwardly, obtaining a planar black hole in such case, as anticipated. 

\vspace{0.3cm}
The three, new, hairy black holes \eqref{eq:nil_metric}, \eqref{eq:solv_metric} and \eqref{eq:sl2_metric}, have horizons parameterized by non-trivial Thurston geometries, that depend on the integration constant $\mu$. Regardless the precise form of the geometry of the base manifold, all of these new solutions are asymptotically locally AdS$_5$, namely
\begin{equation}
R^{\mu\nu}_{\ \ \lambda \rho}=-\frac{1}{\ell^2}\delta^{\mu\nu}_{\lambda \rho}+\mathcal{O}(r^{-2})\ ,
\end{equation}
with $\ell^2=4\alpha$. Note that when one of the cyclic coordinates of the base manifold is made periodic, the integration constant $\mu$ cannot be scaled away. The same situation occurs for Schwarschild-AdS with a planar horizon \cite{Horowitz:1998ha}. It is known that Thurston geometries have constant curvature invariants, and in our case, for the three geometries, the Ricci scalar is given by $-2\mu$. When $\mu=0$, the horizons become planar and one recovers the hairy solutions reported in \cite{Correa:2013bza}. As shown below, the mass of these spacetimes vanishes, but only the integration constant $A$ can be interpreted as hair associated to the scalar field, of primary type, since different values of the integration constant $\mu$ deform the geometry of the conformal boundary. When $A$ vanishes the solution reduces to the topological version of the dimensionally continued black hole reported in \cite{Banados:1993ur}. It is interesting to notice that in such case, the parameter $\mu$ gives a non-vanishing mass that can be written in terms of the volume of the base manifold as well as on its Yamabe functional as shown in equation (85) of reference \cite{Dotti:2007az}. The temperature of these black holes can be computed unambiguously, and is related to the parameter $\mu$ and the horizon radius $r_+$ by
\begin{equation}
\mu=\frac{r_+}{4\alpha}\left(3r_+-16\pi\alpha T\right)\ .
\end{equation}
Considering this relation, one can parameterize the intrinsic horizon geometry in terms of both the temperature $T$ and the horizon radius $r_+$. Interestingly enough, on the line $3r_+=16\pi\alpha T$ the horizon geometry becomes maximally symmetric, and it is invariant under $ISO(3)$, while for a fixed value of $r_+$, and for temperatures other than $\frac{3r_+}{16\pi}$, there is a symmetry breaking on the horizon geometry, since Thurston spaces are homogeneous, but not maximally symmetric.
\section{Thermodynamics \label{sec:Thermo}}

This section conducts a thermodynamic analysis of the solutions with primary hair constructed above. In gravity, the partition function for a thermodynamic ensemble is identified with the Euclidean functional integral in the saddle-point approximation around the classical solution \cite{Gibbons:1976ue}. We require a boundary term $B_E$ whose variation cancels out all the contributions coming from variations of the bulk action, defining a well-posed variational principle \cite{Regge:1974zd}. We show in detail the computation of this boundary term for the Solv-geometry case (the other cases follows similarly), considering the following class of Euclidean metrics
\begin{align}
    ds^2 = N(r)^2 f(r)dt^2 + \dfrac{dr^2}{f(r)} + r^2e^{2x_3}dx_1^2 + r^2e^{-2x_3}dx_2^2 + R(r)^2 dx_3^3 \,.\label{metricminisup}
\end{align}
Notice that the solution \eqref{eq:solv_metric}, can be obtained by the identification $(x_1,x_2,x_3)\rightarrow (x,y,\sqrt{\mu}z)$. In the Euclidean continuation of black holes, regularity requires the Euclidean time to be periodic with period $\beta$, naturally leading to a finite temperature $T$ given by
\begin{align}
    \beta = \dfrac{1}{T} = \dfrac{4\pi}{N(r_{+})f'(r_{+})}\,,
\end{align}
where $r_{+}$ denotes the location of the event horizon.

The evaluation of the action in the Euclidean minisuperspace defined by the metric \eqref{metricminisup} and scalar field $\phi(r)$, in a reduced Hamiltonian form leads to
\begin{align}
    I[g_{\mu\nu},\phi] &= \int d^5x \sqrt{-g} N\mathcal{H} + B_E\,,\\
    &= \beta \sigma \int dr N \mathcal{H} + B_E\,,
\end{align}
where $\sigma$ stands for the volume of the $(t,r)$ constant Euclidean surfaces and
{\small{
\begin{align}
\begin{split}
\mathcal{H} &= \dfrac{1}{32R^3}\Bigg\{ \Big[ (-6\phi^2 r^2 f - 128 \alpha f^2 + 32 r^2 f) R^3 + 128 \alpha r^2 R f \Big] R'' - 256 \alpha f r^2 R'^2 \\
&+ \Big[ (-12 \phi^2 r f + 64 r f - 12 \phi \phi' r^2 f - 3 \phi^2 r^2 f' + 16 r^2 f' - 192 \alpha f f' ) R^3 + (512 \alpha r f + 64 \alpha r^2 f' ) R \Big] R' \\
&+ \Big[ -12 \phi \phi'' r^2 f - 6 \phi \phi' r^2 f' - 6 \phi^2 r f' - 24 \phi \phi' r f + 4 \phi'^2 r^2 f + 32 \nu \phi^4 r^2 + 32 \Lambda r^2 + 32 r f' + 32 f - 6 \phi^2 f \Big] R^4 \\
&+ \Big[ 32 r^2 - 256 \alpha f - 6 \phi^2 r^2 - 128 \alpha r f' \Big] R^2 \Bigg\}\,.
\end{split}
\end{align}
}
From this expression, verifying that the field equations of this reduced action agree with those from the covariant formulation \eqref{eq:feq} and \eqref{eq:kleingordon}, is a straightforward computation. On the minisuperspace \eqref{metricminisup}, the solution reads
\begin{align}
\label{eq:minisup_sol}
N(r)=C\,,\quad f(r)= \dfrac{r^2}{4\alpha} - \mu - \dfrac{3\kappa A^2}{64\alpha r}\,,\quad \phi(r)=\dfrac{A}{r^{3/2}}\,,\quad R(r)= \dfrac{r}{\sqrt{\mu}}\,.
\end{align}
Without loss of generality, we can set $N(r)=1$, and the extrema condition $\delta I = 0$ leads to the following variation for the boundary term {\small{
\begin{align}
\begin{split}
    \delta B_E &= -\dfrac{1}{32R^3}\Bigg\{ \Big[ \Big( (16r^2-192\alpha f - 3\phi^2r^2)R^3 + 64\alpha r^2R\Big) R' + (32r-6\phi^2 r - 6\phi\phi' r^2 ) R^4 -128\alpha rR^2 \Big] \delta f \\
    &+ \Big[20\phi' r^2f + 6\phi r^2f'  \Big]R^4\delta \phi - 12\phi r^2f R^4 \delta \phi' + \Big[-256\alpha r^2 f R' + (-16r^2f' + 64\alpha ff' + 3\phi^2r^2f')R^3 \\ &+ (256\alpha rf - 64\alpha r^2 f') R \Big]\delta R + \Big[(- 6\phi^2 r^2 f-128\alpha f^2 + 32r^2 f) R^3+128\alpha r^2 f R \Big] \delta R' \Bigg\}\,,\end{split}
\end{align}
} where this variation is evaluated at the horizon and infinity. However, on-shell one has
\begin{align}
\begin{split}
    &\delta f = -\delta \mu - \dfrac{3A}{32\alpha r} \delta A\,, \\
    &\delta \phi = \dfrac{\delta A}{r^{3/2}}\,,\quad \delta \phi' = -\dfrac{3\delta A}{2r^{5/2}}\,,\\
    &\delta R = -\dfrac{r\delta \mu}{2\mu^{3/2}}\,,\quad \delta R' = -\dfrac{\delta \mu}{2\mu^{3/2}}\,,
\end{split}
\end{align}
leading to a variation of the boundary term that vanishes identically
\begin{align}
    \delta B_E \equiv 0\,.
\end{align}
Hence, one is obligated to assign vanishing mass and entropy to these new solutions i.e.,
\begin{align}
    \mathcal{M} \equiv 0\,,\quad \mathcal{S} \equiv 0\,.
\end{align}
Of course, one could also assign a constant value to these thermodynamic quantities, nevertheless, requiring the background metric defined by $\mu=A=0$ to have vanishing energy and entropy, would lead to the vanishing of such quantities for the whole family of hairy solutions. 
We must mention that the situation of finding non-trivial solutions in higher curvature gravity, with vanishing mass is not new, as it has been seen for example for BTZ like solutions in Lovelock theories \cite{Anabalon:2011bw} with horizons modeled by products of Thurston geometries, as well as in quadratic gravity with anisotropic backgrounds \cite{Cai:2009ac}. Even more, it was recently shown in \cite{Erices:2024iah} that a stealth scalar field can suppress the mass of the black hole in a more general setting. In our case, the fact that the solution has zero mass is consistent with the fact that in the $\mu\to 0$ case, it was already shown that the mass vanishes \cite{Correa:2013bza}.
\section{Concluding remarks \label{sec:Outro}}

In this paper, we have uncovered a new family of asymptotically locally AdS$_5$ black hole solutions of Einstein-Gauss-Bonnet gravity, characterized by a fully backreacting scalar field of the primary kind. That is to say, the metric exhibits two independent integration constants, one with a gravitational origin, as well as a contribution coming from the backreacting scalar field. When the scalar hair vanishes, the metric still describes a black hole with the horizon given by non-trivial Thurston geometries, which are homogeneous but do not possess the maximum number of Killing vectors. The model considered in this work contains second-order equations for the metric and the scalar field, in contrast with other primary scalar hair solutions \cite{Bakopoulos:2023fmv, Baake:2023zsq} within the realm of Beyond Horndeski theories, which where also generalized in \cite{unouno} for the general case in which the $G_2$ and $G_4$ functions are linearly dependent, analytic, homogenous factors, even identifying a sub-family of theories on which the Weak Energy Condition is fulfilled and the primary hair is conserved due to the shift-symmetry of the identified theory (see also \cite{dosdos} for the thermodynamics of these solutions, using the Euclidean action). The presence of Thurston geometries as solutions of three-dimensional higher curvature theories has already been explored in \cite{Chernicoff:2018hpb,Flores-Alfonso:2021opl,Flores-Alfonso:2023fkd}, while five-dimensional, charged black holes with Thurston horizons have been constructed in \cite{Arias:2017yqj,Bravo-Gaete:2017nkp,minore}, and in \cite{trestres} for Harava-Lifshitz theory while in \cite{cuatrocuatro} for Einstein-Dilaton-Maxwell both on the String and Einstein frames. The solutions with non-trivial Thurston horizons also exist beyond staticity. Indeed, slowly rotating black holes were recently constructed in \cite{Figueroa:2021apr,Figueroa:2024utw}, focusing on the asymptotic symmetry algebra accommodating these solutions in the GR+$\Lambda$ system. Black hole solutions with finite rotation can be constructed in the context of $\mathcal{N}=2$, $D=5$ gauged supergravity \cite{Faedo:2022hle}, as well.

It is well-known that the solutions in vacuum at the Chern-Simons point of the Einstein-Gauss-Bonnet gravity may possess some degeneracy (see e.g. \cite{Dotti:2006cp,Dotti:2007az,Dehghani:2010kd,Cvetic:2016sow,Oliva:2012ff}), in the sense that some of the metric functions are not determined by the field equations, which is a natural feature in dimension five due to the extra gauge invariance that the theory possesses \cite{Hassaine:2016amq}. This can even occur for some of the Thurston horizons as shown in \cite{Peng1,Peng2}. The presence of the scalar field tends to reduce such degenerate behavior. The backreaction of the matter field impacts both on the blackening factors as well as on the horizon geometry.

The thermodynamic properties of the primary hair solution were studied in Section \ref{sec:Thermo}, and we have obtained that both the black hole mass and entropy vanish. In this regard, we proceeded \`a la Regge-Teitelboim, within a minisuperspace, but the same results are obtained using the Wald formalism \cite{Wald:1993nt, Iyer:1994ys}. In spite of this result, our black holes possess a single hair denoted by $A$ coming from the scalar field, since the integration constant $\mu$ that appears in the horizon geometry is inherited by the conformal boundary. 

The existence of the solutions presented in this paper crucially depends on the values of the couplings in the action \eqref{eq:action}. As mentioned above, the part of the action that is quadratic in the scalar field corresponds to that of a conformally coupled scalar, nevertheless, the self-interaction being $\phi^4$ in  dimension five breaks such invariance of the matter sector. Extending these results to higher dimensions in an exhaustive manner is an on-going project, which requires as well to properly identify the family of Thurston geometries we will work with, which is under less control in higher dimensions \cite{Hervik:2003vx,Hervik:2007zz}. In spite of this, let us mention that keeping fixed the non-conformal self-interaction $\phi^4$, we have been able to construct hairy black hole solutions in dimension six modeled by Solv-4 geometries. The solution is given by
{\small{
\begin{equation}
ds^2= -\left(\dfrac{r^2}{12\alpha} - \mu - \dfrac{5\kappa A^2}{288\alpha r^3}\right)dt^2 + \dfrac{dr^2}{\frac{r^2}{12\alpha} - \mu - \frac{5\kappa A^2}{288\alpha r^3}} +r^2\left( e^{2\sqrt{\mu}x_4}dx_1^2 + e^{-\sqrt{\mu}x_4}dx_2^2 +e^{-\sqrt{\mu}x_4}dx_3^2 + dx_4^2\right)\,,
\end{equation}
}
and
\begin{equation}
\phi(r)= \dfrac{A}{r^{5/2}}\,,
\end{equation}
which extremizes the following action
\begin{align}
\label{eq:actiond6}
    S[g_{\mu\nu}, \phi] = \int d^6x\sqrt{-g}\left[ \dfrac{1}{2\kappa}\left( R - 2\Lambda + \alpha \mathcal{G} \right)- \dfrac{1}{2}\nabla_{\mu}\phi\nabla^{\mu}\phi - \dfrac{5}{24}R\phi^2 - \nu \phi^4 \right]\,.
\end{align}
In this case the quadratic part of the action is that of the conformally coupled scalar field but in dimension seven. The couplings fulfill
\begin{align}
    \Lambda = -\dfrac{5}{12\alpha}\,,\quad \nu = -\dfrac{125\kappa}{4608\alpha}\,.
\end{align} This solution reduces to that of \cite{Correa:2013bza} when $\mu\to 0$. We will provide further details of this solution as well as exhaustive generalizations in a future work.

\section*{Acknowledgements \centering}
 
We thank Adolfo Cisterna and Mokhtar Hassaine for enlightening comments. L.G expresses his gratitude to Instituto de Matem\'aticas (INSTMAT) of Universidad de Talca, for its hospitality during the preparation of this manuscript. J.O. is partially supported by FONDECYT Grant 1221504.

\end{document}